# Are We Alone in the Multiverse?

Pushkar Ganesh Vaidya
Indian Astrobiology Research Centre (IARC), Mumbai, India.
pushkar@astrobiology.co.in

**Abstract**

It has been long proposed that black hole singularities bounce to deliver daughter universes. Here the consequences of such a scenario are explored in light of Lee Smolin's hypothesis of Cosmological Natural Selection and Weak Anthropic Principle. The explorations lead towards the answer to the question Are We Alone in the Multiverse?

**Keywords**

Black holes, Cosmic Contact Censorship, Cosmological Natural Selection, Values of the parameters, Weak Anthropic Principle

**Fine Tuning**

There is a general agreement in the scientific community that the values of the parameters are indeed 'fine-tuned' for life, as we know it. [1] Often this is found to be amazing.

The parameters in question are the twenty free parameters of the standard model of elementary particle physics and other fifteen vital cosmological parameters like expansion rate, deceleration parameter, mean mass density, cosmological constant, inflation charge, kind and percentage of nonbaryonic matter and so on. Henceforth these parameters are denoted by $p$. The explanations offered include pure chance, pure chance in a multiverse (Weak Anthropic Principle), by design, how else would we exist (Strong Anthropic Principle), no other values possible and that the values were result of a Cosmological Natural Selection. [2] [3]

**Cosmological Natural Selection**

One attempt to explain $p$ comes from the Smolin's hypothesis of Cosmological Natural Selection. It states that the parameters have the values we observe, because these make the formation of black holes much more likely than most other values.

The collapse of matter into black holes does not end in a nowhere-to-go singularity but results in the formation of a new universe or daughter universe. The $p$ change a little for each daughter universe. Thus, our universe (and others of the multiverse) is a product of mutation analogous to the Darwinian evolution of species, where universes with more black holes have more offspring and dominate the multiverse ensemble. Further, the parameters which optimize black hole formation also allow formation of life as we know it. Thus, our life-permitting universe is a typical member of the multiverse. [2] [3] [4]

There is also a pivotal auxiliary assumption associated with the hypothesis of Cosmological Natural Selection, it states, our universe is typical, i.e. nearly optimal for black hole formation. Also the proposal to test the hypothesis of Cosmological Natural Selection treats the suitability of our universe for life and suitability for production of maximum number of black holes as independent factors. [1]

**Degree of Deviation**

Our universe evidently has $p$ required for life to appear and evolve. However we do not exactly know how much deviation from $p$ is significant and of which parameters, to make a universe unsuitable for life.

If it turns out that life as we know it is only possible with $p$ we observe in our universe, then any deviation from $p$ would result in a universe being unsuitable for life. From this I can deduce that the progeny of our universe cannot be suitable for life. This is because as per the Cosmological Natural Selection $p$ of the progeny will be slightly changed. Hence, a universe suitable for life will never deliver universes suitable for life.

Now, of the progeny, some universes might deliver universes suitable for life but the majority would deliver universes unsuitable for life as $p$ will again be reprocessed. Thus most of the universes of the multiverse would be unsuitable for life. This would mean that the Cosmological Natural Selection hypothesis instead of making our universe typical would make it atypical in the multiverse ensemble.

**WAP Revisited**

According to the Weak Anthropic Principle (WAP) there is an ensemble of universes each with its value of parameters. Some of these universes have values suitable for life and at least in one universe, that's ours, life does exist. Therefore, our universe is atypical. [5]

However, if black holes bounce in such a way that $p$ are not altered then we have a scenario in which our universe will give rise to daughter universes each an exact copy of ours (our universe might be such a copy) in terms of $p$. Thus, even a WAP based beginning will eventually make our kind of universe rather common if not dominant in the multiverse ensemble.

**Where Are They?**

If our universe is typical then it must be surrounded by a population of life-permitting universes like ours. It is probable that each of these should at least harbor one highly technologically advanced civilization. Given enough time it might be possible for such a civilization to travel from one universe to another although for us today it seems impossible even in theory.

So I ask, reminiscent of Enrico Fermi's quip with regards to yet to be found extraterrestrial civilizations in our galaxy, If the multiverse is teeming with technologically advanced civilizations, where are they?

The Rare Earth hypothesis strongly suggests that advanced life (animals) is exceedingly rare in the universe. [6] Thus we might not have much company in our universe but plenty in the multiverse ensemble. If interuniversal travel is possible then we might encounter beings from another universe rather from within our universe.

So can we really expect a visit from a being from another universe?  Well, there is one hurdle and that is Cosmic Contact Censorship.

Cosmic Contact Censorship  tells us that the very characteristics of our universe like vastness, speed of light, effect of inverse square law on EM signal propagation are not suitable for reasonable intergalactic and even interstellar, commuting and communicating. [7]

As all the life-permitting universes in the multiverse ensemble would have the same characteristics, Cosmic Contact Censorship would hold and perhaps more severely as we are talking of interaction between different universes.

## Conclusions

There are two key factors which would eventually answer the question Are We Alone in the Multiverse? First, the life-permitting range of $p$, if there is one. Second, the degree of deviation of $p$ when a daughter universe is formed, if any.

Also, it remains to be seen if any technologically advanced civilization in the multiverse can break the apparent Cosmic Contact Censorship and achieve interuniversal travel. In here, perhaps, also lurks, the hope to prove the existence of the multiverse itself.

## References

[1] How bio-friendly is the universe? - P.C.W. Davies - arXiv: astro-ph/0403050 2 Mar 2004
[2] Is there a Darwinian Evolution of the Cosmos? Some Comments on Lee Smolin's Theory of the Origin of Universes by Means of Natural Selection - Ru diger Vaas. http://arxiv.org/ftp/gr-qc/papers/0205/0205119.pdf
[3] Scientific alternatives to the anthropic principle – Lee Smolin - arXiv: hep-th/0407213 v3 29 Jul 2004
[4] The status of cosmological natural selection - Lee Smolin - arXiv: hep-th/0612185v1 18 Dec 2006
[5] A Critique of Cosmological Natural Selection - Gordon McCabe - http://philsci-archive.pitt.edu/archive/00001648/01/NaturalSelection.pdf
[6] http://www.astro.washington.edu/rareearth/aboutthebook.html
[7] Cosmic Contact Censorship - Pushkar Ganesh Vaidya - arXiv: physics/0702172